
\input harvmac.tex

\def\ra{\rangle}

\def\pb{\overline{\partial}}

\def\np{Nucl. Phys. }
\def\cmp{Comm. Math. Phys. }
\def\pl{Phys. Lett. }
\def\prl{Phys. Rev. Lett. }

\def\pb{\overline{\partial}}

\def \rb{\bar{r}}

\hfill{SWA /93-94/15 , WIS-93/1-CS , RI-93/69}
\Title{}{\vbox{\centerline {A Class of String Backgrounds as a   }
\vskip2pt\centerline{Semiclassical Limit of WZW Models }}}
\smallskip
\centerline{\it
D.I. Olive       }
\centerline{ Department of Mathematics, University College of Swansea
                                  }
\centerline{ Swansea SA2 8PP, WALES UK                }
\smallskip
\centerline{\it E. Rabinovici }
\centerline{Racah Institute of Physics,Hebrew University}
\centerline{Jerusalem, ISRAEL}
\smallskip
\centerline{and}
\centerline{\it A.   Schwimmer}
\centerline{SISSA and INFN, Trieste 34013, ITALY}
\centerline{and Weizmann Institute, Rehovot 76100, ISRAEL}
\noindent
\vskip 1cm
\baselineskip 18pt
\noindent
  A class of string backgrounds associated with non semi-simple groups
 is obtained as a special large level limit of ordinary WZW models.
  The models have an integer Virasoro central charge and they include
  the background recently studied by Nappi and Witten.
\Date{November 1993}

\magnification=1200 \hsize=6.2 truein \vsize=8.9 truein

In this note we discuss from an algebraic point of view a class of exact
conformal field theories
which are based on WZW models \ref\witten {E.Witten , \cmp 92(1984) 455}\
on  group manifolds which are neither compact
nor semi-simple. The feature of these models is that the
Virasoro central charge  is independent of the levels of the
affine Lie algebra
and is thus an integer equal to the dimension of the group manifold.

A non-linear sigma model Lagrangian representation is formally available for
all WZW models. To a
given group $G$ corresponds an appropriate target space metric and torsion in
the non-linear sigma
model. In that setting the criteria  for conformal invariance is the exact
vanishing of the beta
function of the theory. Usually it is difficult to verify such a condition in
the absence of some
underlying algebraic structure. For non-abelian groups neither the metric nor
the torsion are
flat but they are nevertheless related in a manner reflecting the
parallelisability of the
manifold \ref\zachos{T.Curtright and C.Zachos, \prl 53 (1984) 1799 }\ .
 The Virasoro central charge is not necessarily integral and this can
be understood in a
\lq\lq free field representation" as arising from the necessity to project out
states of the free
fields by a Feigin-Fuks like construction.

Recently \ref\nappi{C.R.Nappi and E.Witten , preprint
IASSNS-HEP-93/61  }\
it has been shown that a WZW model based on a central extension of
the Euclidean  group
in two dimensions has a central charge equal to four,
independently of the
level. This result was
established by considering the beta functions in a non-linear sigma model
description with the
central charge four found in perturbation theory. Non perturbative arguments
were also given.
Reinterpreting the algebra as a Wigner contraction of an $so(3)\times u(1)$
algebra we are able to
present a large  class of models
 based on a certain contraction of the group
$G\times H$ where the
latter is a compact subgroup of the former, assumed compact and simple.
  For the affine Lie algebra the construction involves a correlated large
 $x$ limit  ($x$ being the level)
 of $G$ and $H$. Therefore the string backgrounds represented
 by the WZW models based on the centrally extended non semisimple Lie
 algebras are actually semiclassical limits of the appropriate simple
 group WZW models.
As a result the
Virasoro central charge
equals the total dimension of the group (which is unchanged in the contraction)
and the metric is
nonsingular with the number of minus eigenvalues equal to the dimension of $H$.

 The Conformal Theories corresponding to the $G\times H$  WZW models
 for compact
$G$ and $H$  , are exactly solvable for arbitrary positive integer
levels. The semiclassical limit we are taking requires negative
 integer levels for the group $H$. In the conventional large (positive) $x$
limit
                                          the remnant of
  a compact WZW model is described in the leading order
                  by the modes  of a  number of free bosons equal to
  the dimension of the group.
                             The limit we are taking
        is different, requiring another type of scaling behaviour
 for both the positive and negative
 level components.
 We expect           that the solution of this class of
 backgrounds will be a deformation of the theory having $dim G$ positive
  signature bosons and $dim H$ negative signature ones.

Since the other known examples of integer central Virasoro charge are of
physical significance we
hope the present examples will prove of interest as possible string
backgrounds. Thus it has been
known for some time that the central charge of level one WZW models for simply
laced Lie groups
equals the rank of the group,
$r$. In that case it is understood from an algebraic point of view that the non
flat metric arising
in the nonlinear sigma model can be replaced by the field theory of $r$ free
bosons on the torus
$U(1)^r$ which can be conformally embedded in the level one WZW model. From a
string point of view
the lesson learnt is that, at the Planck scale, the geometrical picture and
even the dimension are
ambiguous and may afford several complementary descriptions.
 Calabi-Yau spaces \ref\candelas {P.Candelas , G.Horowitz ,
 A.Strominger  and E.Witten \np B258 (1985) 46}\
 and monopole backgrounds \ref\banks
  {T.Banks, M.Dine, H.Dijkstra and W.Fischler, \pl B212(1988) 45}\
serve as further
examples of non flat spaces with integral Virasoro central charge but here the
algebraic understanding of this fact remains elusive.
  Gravitational shock waves \ref\amati
  {D.Amati and C.Klimcik , \pl B219(1989) 443 }\ have also been shown
  to provide exact string backgrounds with integrally valued central
  charge.

\def\ui{u^i} \def\uj{u^j} \def\uk{u^k}  \def\vi{v^i}
\def\vj{v^j} \def\vk{v^k}
\def\mab{iM_{\alpha\beta}^i} \def\sabc{s^{\alpha\beta\gamma}}
\def\ra{R^{\alpha}} \def\rb{R^{\beta}}
\def\rc{R^{\gamma}}
\def\ep{\epsilon} \def\ti{T^i} \def\tj{T^j} \def\tk{T^k}  

\def\pa{P^{\alpha}} \def\pb{P^{\beta}} \def\de{\delta_{m+n,0}}
We consider a compact, simple Lie group $G$ containing a compact subgroup
isomorphic to $H$. We
choose orthonormal bases for the hermitian generators of their Lie algebras
$g=\{\ui, \ra\}$ and
$h=\{\vi\}$ so that the commutation relations read:

$$[\ui,\uj]=if^{ijk}\uk, \qquad [\vi,\vj]=if^{ijk}\vk,\eqno(1a)$$
$$[\ui,\ra]=\mab\rb,\eqno(1b)$$
$$[\ra,\rb]=\mab\ui+i\sabc\rc,\eqno(1c)$$
where $f^{ijk}$ and $\sabc$ are totally antisymmetric and $M^i$ an
antisymmetric matrix.
We are going to consider a Wigner contraction of $G\times H$ by defining
$$\ti=\ui+\vi, \qquad F^i={\ep\over2}(\ui-\vi),\qquad
\pa=\sqrt{\ep}\ra,\eqno(2)$$
where $\ep$ is a parameter that will be taken to zero. In the limit we find:
$$[\ti,\tj]=if^{ijk}\tk,\qquad [\ti,\pa]=\mab\pb,\eqno(3a)$$
$$[\ti,F^j]=if^{ijk}F^k, \qquad [\pa,\pb]=\mab F^i,\eqno(3b)$$
$$[\pa,F^i]=0,\qquad [F^i,F^j]=0.\eqno(3c)$$
These relations define the algebra $m$ which is the contraction of
$g\oplus h$. Since the $\{F^i\}$
generate an invariant abelian subalgebra $m$ is not semisimple. It
possesses a gradation $m=m_{(0)}+m_{(1)}+m_{(2)}$ with the scaling
symmetry $m_{(n)}\rightarrow\lambda^n m_{(n)}$ reflecting the
disappearance of the parameter $\ep$. Here $m_{(0)}=\{T^i\}$,
$m_{(1)}=\{\pa\}$, and $m_{(2)}=\{F^i\}$. Notice that although the
structure constant $\sabc$ has dropped out of $m$, there is no need for
the symmetric space structure for $G/H$ that would be implied if it vanished.

Supposing now, for definiteness, that $h$ is either simple or one
dimensional, we see that $g\oplus h$ has two independent quadratic Casimirs,
namely those of $g$
and $h$, and we shall see that the contraction, $m$, inherits two Casimirs in
different form.
Substituting (2), the two Casimirs of $g$ and $h$ become

$$C(g)\equiv\sum (\ui)^2+\sum(\ra)^2 ={1\over\ep^2}\sum
(F^i)^2+{1\over\ep}(\sum (\pa)^2+\sum\ti
F^i)+{1\over4}\sum(\ti)^2,$$
$$C(h)\equiv\sum (\vi)^2={1\over\ep^2}\sum (F^i)^2-\sum\ti
F^i+{1\over4}\sum(\ti)^2.$$

The leading term, $$C_1=\sum (F^i)^2,\eqno(4a)$$
 as $\ep$ tends to zero is common to both these
Casimirs, and is indeed itself a Casimir of the contracted algebra $m$. The
leading term after its
elimination, namely
$C(g)-C(h)={1\over\ep}(\sum(\pa)^2+2\sum\ti F^i),$
yields a second Casimir,
$$C_2=\sum(\pa)^2+2\sum\ti F^i,\eqno(4b)$$
 which, unlike the first, corresponds to a
nonsingular quadratic form. We see from the first expression for $C_2$ that its
metric possesses
$d(g)$ positive signs and
$d(h)$ negative signs ($(d(g)$ denotes the dimension of $g$).

Thus the simplest example, $g=so(3)$ and $h=u(1)$, yields a four dimensional
algebra with one minus
sign in the metric. This is the candidate for the space-time background
considered in \nappi .

Now let us affinise $g\oplus h$ in order to obtain the affine Lie       algebra
$\hat g \oplus\hat
h$ with two levels, $x(g)$ and $x(h)$, which are expected to be integers of the
same sign as the
corresponding Virasoro $L_0$. After the substitution (2) applied to the graded
counterparts of (1)
we obtain
$$[\ti_m,\tj_n]=if^{ijk}\tk_{m+n}+(x(g)+x(h))m\delta_{m+n,0},\eqno(5a)$$
$$[\ti_m,F^j_n]=if^{ijk}F^k_{m+n}+(x(g)-x(h)){\ep\over2}m\de,\eqno(5b)$$
$$[\pa_m,\pb_n]=\mab F^i_{m+n}+x(g)\ep m\de.\eqno(5c)$$
The central terms $x(g)$ and $x(h)$ do not appear in the other three
commutators which we have
accordingly omitted. We want the new central terms to be finite as
$\ep\rightarrow0$, that is we
want $x(g)+x(h)$, $\ep x(g)$ and $\ep(x(g)-x(h))$ to be finite and this means
that the levels $x(g)$ and $x(h)$ must become indefinitely large in correlated
ways:

$$x(g)={\alpha\over\ep}+\delta,\qquad
x(h)=-{\alpha\over\ep}+\beta-\delta.\eqno(6)$$
So
$$x(g)+x(h)\rightarrow\beta,\qquad \ep x(g)\rightarrow\alpha
\quad\hbox{and}\quad
\ep(x(g)-x(h))\rightarrow\alpha.\eqno(7)$$
Thus $x(g)$ and $x(h)$ must have opposite signs and the affinisation of $m$,
$\hat m$, must possess
two central terms $\alpha$ and $\beta$ which are no longer integers. That there
are two levels is
to be expected as $m$ has two invariant bilinear forms just as it has two
quadratic Casimirs.

Now consider the Sugawara construction for $\hat g\oplus \hat h$. Given the two
levels $x(g)$ and
$x(h)$ it is unique and has the Virasoro c-number
$$c={x(g)d(g)\over x(g)+\tilde h(g)}+{x(h)d(h)\over x(h)+\tilde
h(h)}\rightarrow d(g)+d(h),
\eqno(8)$$ in the limit $\ep\rightarrow 0$. A key    observation is that this
limit is independent
of $\alpha$ and $\beta$, being simply the dimension of $m$, which is, of
course, an integer.

Our final step is to check how the Sugawara construction for $g\oplus h$
reduces, via the Wigner
contraction, to the Sugawara construction for $m$. As the latter is not
semisimple there will be
novel features. We start with the unique construction

$$L={:u^2:+:R^2:\over2(x(g)+\tilde h(g))}+{:v^2:\over2(x(h)+\tilde
h(h))},\eqno(9)$$
and substitute for $\hat m$ via (2) and (7), finding

$$={:P^2:+2:FT:\over2\alpha}-{(\tilde h(g)+\tilde
h(h)+\beta):F^2:\over2\alpha^2},\eqno(10)$$
where we have used a natural notation. Notice that this involves the quadratic
Casimir expressions
$C_1$ and $C_2$, (4) and the levels $\alpha$ and $\beta$, relevant to $\hat m$.
In particular the
constant $\delta$ has disappeared.

 The gradation of the Lie algebra mentioned above has its counterpart for the
Kac-Moody algebra provided we scale  $\alpha$ as $\lambda^2$ and we leave
$\beta $ unchanged.The Virasoro generators are invariant under the rescaling.

   We are going to show now  that the contraction can be done directly
 for the WZW model. In this way, we shall find the
 action which corresponds to the algebra
 (5) together with the Sugawara construction (10).

    A WZW action at a   conformal point at level x, for an
     arbitrary group G
   has the form:
         $$ S=x \int W(G) \eqno(11)$$
    where the integration is over two dimensional
    space time and $W(G)$
   is really a functional of the Cartan form $\gamma^{-1}d\gamma$
   where $\gamma$ now denotes an element of the Lie group.
   The Cartan
    form , being an element in the Lie algebra , can be expanded
     in the generators of the Lie algebra, $Z^i$ :
      $$  \gamma^{-1}d\gamma=z_i Z^i \eqno(12)$$
      Then the functional $W   $ is given by:
      $$ W   = c^{ij}z_iz_j +       c^{km}\kappa^{ij}_m z_iz_jz_k \eqno(13)$$
       where $c^{ij}$ is an invariant, invertable
       metric on the algebra,
     $\kappa ^{ij}_k$ are the structure constants of the group, and a
     numerical factor between the two terms was   omitted.

     We start with the action corresponding to the WZW model for
     $G\times H$
 with levels $x(g)$ and $x(h)$ respectively:

   $$ W=x(g)W(G) +x(h)W(H) \eqno(14)$$
 Since the groups
         G and H are
         now considered to be simple and compact, as before,
                                $c^{ij}$  can be taken to be
    the Cartan metric , diagonal in the basis specified by   (1).
  We call $U_i$ , $r_{\alpha}$ , $V_i$
  the components of the Cartan form along the directions $u$ , $R$
  and $v$ ,                               respectively.

 If the parameters of the group G
 were $\theta_i $ and $ \phi_{\alpha}$ corresponding to $u^i $ and
 $R^{\alpha}$ respectively and those of $H$ , $\chi^i$, one can define
 parameters corresponding to $T$ , $F$ and $P$ by:
   $$ \bar \theta_i={1\over2} (\theta_i+\chi_i),\quad
    \bar\chi_i={1\over{\epsilon}} (\theta_i-\chi_i),\quad
  \bar\phi_{\alpha}={1\over\sqrt{\ep}}\phi_{\alpha} \eqno(15)$$
       From (12) it is clear that, in the contracted limit, the group
   manifold corresponds to one copy of $H$, a tangent plane to it
   at the origin and tangent, non-compact directions corresponding
   to $R$. In the WZW action after the contraction the basic
    fields will be the ones specified by (15).

   The WZW action after the Wigner
   contraction can be obtained easily : using
       (2) , one expands the Cartan form along the directions $T$, $F$
    and $P$. Calling the contracted
    components $t_i$, $f_i$ and $p_{\alpha}$ their relation to the
    original components is given by:
      $$U_i=t_i+{1\over2}  \ep f_i\eqno(16a)$$
      $$V_i=t_i-{1\over2} \ep f_i\eqno(16b)$$
      $$r_{\alpha}=\sqrt{\ep}p_{\alpha}\eqno(16c)$$

   Using   (16) for the components in    (13) , and the expressions (6)
  for the levels , the limit $\ep \rightarrow 0$ is taken. We obtain
  for the Wigner contracted WZW action W :
      $$W=\alpha (2 t_i f^i +p_{\gamma} p^{\gamma}+3f^{ijk}t_it_j f_k)+
          \beta(t_it_i+f^{ijk}t_it_jt_k) \eqno(17)$$
     The interpretation of (17) is clear : it is the WZW action
   corresponding to  the group whose Lie algebra $m$ is
   specified by the commutation relations
     (3) , at level $\alpha$ and with the nonsingular invariant metric:
      $$ c^{TF}=1 ,\qquad c^{PP}=1 ,\qquad c^{TT}=\beta / \alpha\eqno(18)$$
      The inverse of this metric gives a linear combination of
       the Casimirs (4). The Sugawara construction (10) follows
    from (17) when the one loop corrections are included.

      We postpone the detailed study of the limit
      of the  $G\times H$
   theory proposed above and its relation to the solution of the
    non semi-simple models to a further publication. For the time being we
simply remark
    that scaling the    generators of $G$   and  $H$ as $\sqrt{x}$ in the
manner
     corresponding to the ordinary semiclassical limit which leads to
      certain type of free bosons
     is inconsistent
     with (2) and (6).

    The models discussed have moduli associated with the truly
   marginal operators which can be constructed from the currents. These
    may lead to more general backgrounds of the type discussed in
        \amati .

     The work of E.R. was partially supported by the Israel Science
   Foundation, the USA-Israel Binational Science Foundation and by the
   Ambos Monell Foundation, that of A.S. by the USA-Israel Binational
   Science Foundation. One of us (A.S.) would like to thank the Theoretical
Physics
   Group in  Swansea,  where part of this work
   was performed,
   for the hospitality extended to him.

\listrefs

\bye